\documentclass[twoside,twocolumn,english,aps, prl,  superscriptaddress]{revtex4}
\pdfoutput=1
\usepackage[latin9]{inputenc}
\usepackage{graphicx}
\usepackage{amssymb}
\usepackage{esint}

\usepackage{babel}

\begin{document}

\title{Measurement-based Synthesis of multi-qubit Entangled States in Superconducting
Cavity QED}

\author{Ferdinand Helmer}

\affiliation{{\small Department of Physics, CeNS, and ASC, Ludwig-Maximilians-Universität,
Theresienstrasse 37, D-80333 Munich, Germany}}

\author{Florian Marquardt}

\affiliation{{\small Department of Physics, CeNS, and ASC, Ludwig-Maximilians-Universität,
Theresienstrasse 37, D-80333 Munich, Germany}}
\begin{abstract}
Entangled multi-qubit states may be generated through a dispersive
collective QND measurement of superconducting qubits coupled to a
microwave transmission line resonator. Using the quantum trajectory
approach, we analyze the stochastic measurement traces that would
be observed in experiments. We illustrate the synthesis of three-qubit
W- and GHZ-states, and we analyze how the fidelity and the entanglement
evolve in time during the measurement. We discuss the influence of
decoherence and relaxation, as well as of imperfect control over experimental
parameters. We show that the desired states can be generated on timescales
much faster than the qubit decoherence rates. 
\end{abstract}
\maketitle
\global\long\def\ket#1{\left|#1\right\rangle }

\global\long\def\bra#1{\left\langle #1\right|}

\global\long\def\s{\hat{\sigma}}

\def\+{\;} 
\def\-{\!}

\section{Introduction}

The realization of quantum-optical concepts in condensed matter systems
has led to remarkable progress during the past few years. One of the
prime examples is the study of quantum electrodynamics (QED) in superconducting
circuits. Earlier suggestions to implement the Jaynes-Cummings model
in the solid state \citep{2001_Marquardt_CooperBox,2001_BuissonHekking_Cat,2003_YouNori_CircuitQED}
were followed by a proposal \citep{2004_02_BlaisEtAl_CavityProposal}
to employ on-chip microwave resonators and couple them to artifical
atoms in the form of superconducting qubits. This seminal idea was
soon thereafter realized experimentally \citep{2004_09_WallraffEtAl_MicrowaveCavity},
creating a solid-state analogue of conventional optical cavity QED
\citep{2002_11_MabuchiDoherty_CavityQED_SCIENCEReview}. The tight
confinement of the field mode and the large electric dipole moment
of the {}``atom'' yield extraordinary coupling strengths. As a result,
these highly tunable systems have been employed to demonstrate experimentally
a variety of achievements, including: The Jaynes-Cummings model in
the strong-coupling regime \citep{2004_09_WallraffEtAl_MicrowaveCavity,2004_nature_Mooi_Quibit_Resonator,2006_03_Semba_JapaneseCircuitQED},
Rabi and Ramsey oscillations and dispersive qubit readout \citep{2005_04_SchusterWallraff_PRL_DephasingByReadout,2005_08_Wallraff_PRL_UnitVisibility},
generation of single photons \citep{2007_Yale_Single_Microwave_Source_condmat}
and Fock states \citep{hofheinz2008gfs,wang2008mdf}, cavity-mediated
coupling of two qubits \citep{majer2007csq,2007_Simmonds_2qubitscavitynature},
setups with three qubits \citep{2008_Wallraff_3_qubits_Tavis_Cummings},
Berry's phase \citep{leek2007obs}, and the measurement of the photon
number distribution \citep{2007_Yale_Nature_Photon_Number_Splitting}.

\begin{figure}[b]
\includegraphics[width=1\columnwidth]{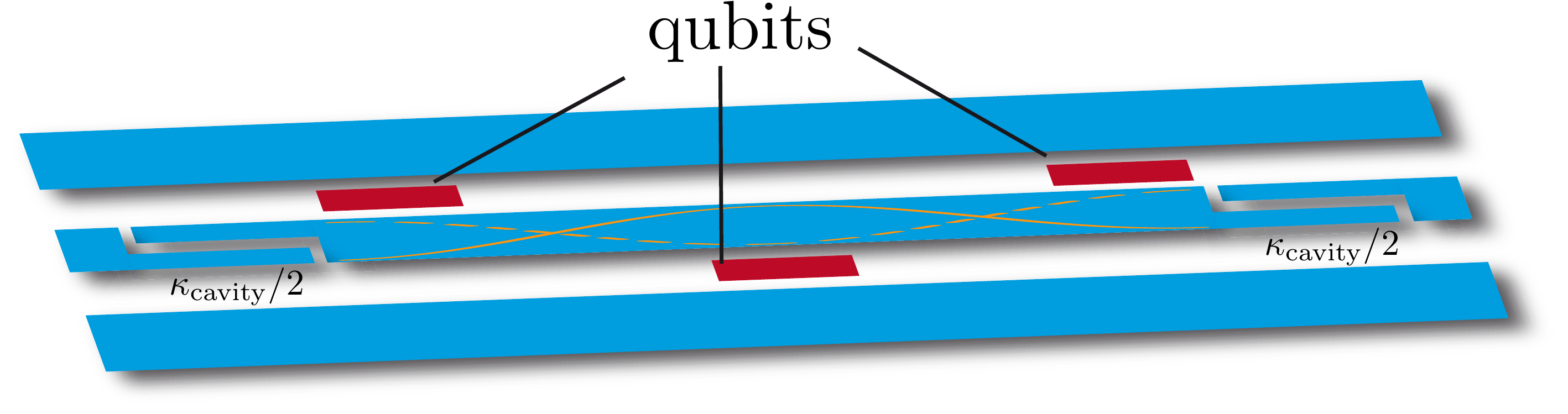}

\caption{(Color online) Schematic setup: Three superconducting qubits (red)
are coupled to a mode of a coplanar microwave resonator (blue). The
measurement of the phase shift of a transmitted microwave beam can
be utilized to rapidly synthesize e.g. maximally entangled multi-qubit
states like GHZ and W-states.}
\label{fig:setup}
\end{figure}

The strong coupling makes dispersive quantum non-demolition (QND)
readout possible, both for qubit states  and for detecting single
photons \citep{Helmer2008Quantum}. QND measurements are ideal projective
measurements that reproduce their outcome when repeated \citep{1980_Science_Braginsky_QND_Measurement,1992_BraginskyKhalili_QuantumMeasurement}.
Any QND measurement may be applied to (probabilistically) generate
states. In particular, having several qubits inside a common cavity
(as realized in recent circuit QED experiments \citep{majer2007csq,2007_Simmonds_2qubitscavitynature},
for a schematic setup see Fig.~(\ref{fig:setup}))), one may produce
entangled multi-qubit states, even without employing directly any
qubit-qubit coupling. In the context of circuit QED, this option has
been investigated previously in a series of remarkable studies \citep{sarovar2005hfm,2008arXiv0812.0218H,rodrigues2008esc}.
However, these consider primarily two qubits, with a recent work \citep{2008arXiv0812.0218H}
discussing the extension to more qubits in general terms. The present
paper aims to go beyond these studies in several aspects. First, we
present necessary conditions for being able to generate arbitrary
multi-qubit states out of a given subspace of the total multi-qubit
Hilbert space, using only single qubit operations and subsequent collective
measurement. Second, we carry out detailed quantum jump trajectory
simulations also for the case of three qubits, where W and GHZ states
may be produced. We show how entanglement is generated in the course
of the measurement process, paying attention to the effects of relaxation
and decoherence. Moreover, we analyze how imprecise fine-tuning of
experimental parameters would lead to a loss of entanglement after
its initial, transient generation. Finally, we comment on possible
experimental realizations. Such a measurement-based scheme complements
other approaches for entanglement-generation in circuit QED \citep{siewert2002idj,2004_02_BlaisEtAl_CavityProposal,lantz2004jjq,wallquist2005sqn,wei2006gac,wallquist2006scs,steffen2006met,maruyama2007eet,you2007eos}
, based on unitary dynamics, and may prove advantageous for some purposes,
since generation and measurement are combined into one step. It might
also be used to generate entanglement between qubits in spatially
separated cavities, without any direct interaction.

\section{model}

We investigate a QND scheme utilizing the coupling of superconducting
qubits to a bosonic field mode of a microwave resonator as examined
in \citep{2004_02_BlaisEtAl_CavityProposal,2004_09_WallraffEtAl_MicrowaveCavity}.The
presence of excitations in the qubits inside the cavity gives rise
to a frequency shift of the cavity mode, which can be observed dispersively
via the phase shift of a transmitted beam. In turn, the measurement
backaction leads to a projection of the qubits on a state that depends
(a) on the chosen set of couplings and (b) the initial (product) state
the qubits are prepared in.

We consider a system of a driven cavity mode coupled to $N$ qubits
\begin{eqnarray}
\hat{H} & = & \hbar\omega\left(\hat{a}^{\dagger}\hat{a}+\frac{1}{2}\right)+\sum_{i=1}^{N}\frac{\epsilon_{i}}{2}\left(\hat{\sigma}_{i}^{z}+1\right)\nonumber \\
 & + & \sum_{i=1}^{N}g_{i}^{0}(\hat{\sigma}_{i}^{+}\hat{a}+\hat{\sigma}_{i}^{-}\hat{a}^{\dagger})+\frac{\kappa_{{\rm cavity}}}{2}\left(\varepsilon\hat{a}^{\dagger}+\hat{a}\varepsilon^{*}\right)\nonumber \\
 & + & \hat{H}_{{\rm {\rm decay}}}\,,\label{eq:Hamiltonian}\end{eqnarray}
a model commonly known as the Tavis-Cummings model which has been
recently realised experimentally for $N=2$ \citep{majer2007csq,2007_Simmonds_2qubitscavitynature}
and $N=3$ \citep{2008_Wallraff_3_qubits_Tavis_Cummings}. The first
term of this Hamiltonian describes the cavity mode with a frequency
$\omega$, the second all qubit energies, the third term realizes
the Jaynes-Cummings coupling for each qubit to the cavity with bare
coupling constants $g_{i}^{0}$, while the last term describes the
driving of the cavity with the readout microwave tone which will yield
$|\varepsilon|^{2}$ photons in the resonator on average ($\kappa_{{\rm cavity}}$
is the intensity decay rate for the cavity).

In the limit where all the qubits are strongly detuned from the cavity,
it is well-known \citep{2004_02_BlaisEtAl_CavityProposal} that the
qubits impart a state-dependent phase shift on the cavity mode and
the effective Hamiltonian can be written as:

\begin{eqnarray}
\hat{H} & = & \hbar\omega\left(\hat{n}+\frac{1}{2}\right)+\sum_{i=1}^{N}\frac{\epsilon_{i}}{2}\left(\hat{\sigma}_{i}^{z}+1\right)\nonumber \\
 & + & \sum_{i=1}^{N}\frac{(g_{i}^{0})^{2}}{\Delta_{i}}\frac{\left(\hat{\sigma}_{i}^{z}+1\right)}{2}\hat{n}+\frac{\kappa_{{\rm cavity}}}{2}\left(\varepsilon\hat{a}^{\dagger}+\hat{a}\varepsilon^{*}\right)+\hat{H}_{{\rm {\rm decay}}}\,\nonumber \\
 & = & \hbar\left[\omega+\sum_{i=1}^{N}\frac{(g_{i}^{0})^{2}}{\Delta_{i}}\frac{\left(\hat{\sigma}_{i}^{z}+1\right)}{2}\right]\left(\hat{n}+\frac{1}{2}\right)+\sum_{i=1}^{N}\frac{\epsilon_{i}}{2}\left(\hat{\sigma}_{i}^{z}+1\right)\nonumber \\
 & + & \frac{\kappa_{{\rm cavity}}}{2}\left(\varepsilon\hat{a}^{\dagger}+\hat{a}\varepsilon^{*}\right)+\hat{H}_{{\rm {\rm decay}}}\,\label{eq:eff_ham}\end{eqnarray}

The Hamiltonian (\ref{eq:Hamiltonian}) also induces an effective
flip-flop interaction \citep{2004_02_BlaisEtAl_CavityProposal,2007_09_Majer_TwoQubits}
of strength $J_{\alpha\beta}=g_{\alpha}g_{\beta}(\Delta_{\alpha}+\Delta_{\beta})/(2\Delta_{\alpha}\Delta_{\beta})$
between each pair of qubits $(\alpha,\beta)$ in the same cavity (for
couplings $g_{\alpha(\beta)}$ and detunings from the cavity $\Delta_{\alpha(\beta)}$,
in the dispersive limit $\left|g\right|\ll\left|\Delta\right|$):

\begin{equation}
\hat{H}_{\alpha\beta}^{{\rm flip-flop}}=J_{\alpha\beta}\left(\hat{\sigma}_{\alpha}^{+}\hat{\sigma}_{\beta}^{-}+{\rm h.c.}\right).\label{Hflipflop}\end{equation}
When simulating the master equation (\ref{eq:masterequation}) to
be derived from the Hamiltonian (\ref{eq:eff_ham}), we neglect this
interaction for several reasons. (i) In a concrete experiment the
qubit energies could always be chosen very different, such that this
unwanted interaction does not play a role, since the qubits are non-resonant.
(ii) As we will argue later in more detail, the measurement rate $\bar{\Gamma}$
is usually much larger than $J_{\alpha\beta}$, thus making the effects
of the interaction negligibly small even when the qubits are in resonance
with each other. (iii) We note that all the states we consider as
examples are eigenstates of the flip-flop interaction, Eq.~(\ref{Hflipflop}).
Therefore, even if the qubits are chosen to be in resonance (as is
ultimately assumed in our simulations), the interaction will not have
any important effect on the dynamics besides trivial phases between
subspaces that are rendered mutually incoherent by the measurement
anyway. Thus, we will neglect the flip-flop interaction.

\section{necessary conditions for the generation of given target states}

$\frac{}{}$The protocol we are envisaging is to first produce a product
state, using local operations on the individual qubits, and then to
project on an entangled state by measurement. This scheme will be
successful with a certain probability. Our aim in the present section
is to briefly discuss the necessary conditions that must be met to
be able to generate a given class of entangled states. It goes without
saying that once an entangled state has been reached, one may then
apply further local operations to reach a corresponding subspace of
the full multi-qubit Hilbert space. 

Let us first fix notation. The coupling strengths $g_{i}$ determine
the phase shifts induced by the individual qubits, 

\begin{equation}
g_{i}\equiv\frac{(g_{i}^{0})^{2}}{\Delta_{i}},\end{equation}
and for convenience we will collect them into the coupling vector

\begin{equation}
\overrightarrow{G}\equiv\left(\frac{g_{i}}{\bar{g}}\right),\, i=1...N\,,\end{equation}
where the overall strength $\bar{g}$ just determines the measurement
time-scale but does not affect the reachable states. 

Using $ $qubit excitation operators $\hat{n}_{i}\equiv\frac{\left(\hat{\sigma}_{i}^{z}+1\right)}{2}$
, we define the measurement operator $\hat{N}$ as 

\begin{equation}
\hat{N}\equiv\sum_{i=1}^{N}\frac{g_{i}}{\bar{g}}\hat{n}_{i}\,.\label{eq:Def_Op_N}\end{equation}
Note that even in current experiments, the $g_{i}$ are tunable in
magnitude and sign simply by choosing the detuning $\Delta$ appropriately.
The frequency shift imposed on the microwave cavity will then be \begin{equation}
\Phi=\sum_{i=0}^{N}\phi_{i}=\bar{g}\langle\hat{N}\rangle=\sum_{i=1}^{N}g_{i}\left\langle \hat{n}_{i}\right\rangle =\sum_{i=1}^{N}g_{i}n_{i\,,}\label{eq:Def_phase_shift}\end{equation}
where we have defined $n_{i}:=\langle\hat{n}_{i}\rangle\in[0,1]$
as the excitation number of the $i$-th qubit.

The desired entangled state $\ket{\Psi}\equiv\sum_{j=1}^{2^{N}}\alpha_{j}\ket{\varphi_{j}}$
has complex amplitudes

\begin{equation}
\overrightarrow{\alpha}\equiv\left(\alpha_{j}\right),\, j=1...2^{N}\,\end{equation}
in the energy eigenbasis of the qubits (a product basis that diagonalizes
$\hat{n}_{i}$). Thus, we have $n_{1,\ket{\Psi}}=\bra{\Psi}\hat{n}_{1}\ket{\Psi}=\sum_{j=1}^{2^{N}}|\alpha_{j}|^{2}\bra{\varphi_{j}}\hat{n}_{1}\ket{\varphi_{j}}$.

In the following we derive \emph{necessary} conditions for being able
to produce arbitrarily chosen states out of some $M$-dimensional
Hilbert space that is spanned by a subset of $M$ basis states $\ket{\varphi_{j}}$.
In order to generate a certain target state given by arbitrary $\overrightarrow{\alpha}$,
we need to adjust the couplings such that all base kets with non-vanishing
$\alpha_{j}$ yield the same phase shift. Assume the amplitude vector
of the target state has $M\in\{1,...,2^{N}\}$ non-zero entries $\alpha_{j},j\in\{1,..,2^{N}\}$
where the corresponding indices can be written as a family $F_{\alpha}$
with $\dim(F_{\alpha})=M$. Then the goal is to use the measurement
to project the system onto the subspace given by ${\rm {\rm span}\left(\left\{ \ket{\varphi_{j}}\left|j\in F_{\alpha}\right.\right\} \right)}.$ 

In the simplest case this is directly possible by choosing (i) an
appropriate initial product state of the  qubits (to fix the amplitudes)
and (ii) a suitable coupling vector (to project onto the correct subspace). 

Choosing an arbitrary initial product state allows for the choice
of $2N$ complex amplitudes. Due to normalization of the $N$ single
qubit states and a an arbitrary global phase for each of those states,
we essentially have $2N$ real parameters to choose.

The amplitude vector of the target state will - up to a constant common
factor due to the renormalization after projection - be determined
by the amplitudes of this initial state. This suggests that, in general
(i.e. for arbitrary target states), we can only aim at reaching states
that satisfy \begin{equation}
2M-2\leq2N.\label{eq:ineq_complex-1}\end{equation}
Again, we had to subtract $2$ to account for the irrelevant global
phase and normalization. 

Note that for the maximal value of $M=2^{N}$, the last inequality
does not hold for any $N>1$ and we recover the fact that arbitrary
states are in general not product states. Note that we have just found
a necessary condition for constructing \emph{arbitrary }states out
of an $M$-dimensional subspace. When choosing particular states,
e.g. trivially separable states, one may still be able to construct
those even if they formally violate $ $Eq.~(\ref{eq:ineq_complex-1})). 

We now turn to the question when it is possible to choose the couplings
such that the measurement cannot distinguish the components of the
target state from each other. This requirement of equal phase-shifts
formally corresponds to a set of $M-1$ equations 

\begin{equation}
\bra{\varphi_{i}}\hat{N}\ket{\varphi_{i}}=\bra{\varphi_{j}}\hat{N}\ket{\varphi_{j}}\label{eq:phase_shift_eq_general}\end{equation}
where i,j denote successive indices out of $F_{\alpha}$. 

As tunable parameters to our disposal we effectively have $N-1$ couplings
(discounting the overall strength $\bar{g}$) so this set of equations
will in general be solvable as long as $M\leq N$ is fulfilled. 

As we will demonstrate below in several examples, some of the most
interesting entangled states, such as W and GHZ states for three qubits
can be synthesized by this scheme. Indeed, they have $M=N$ for the
W- and $M=2$ for the GHZ-states and thus satisfy the necessary conditions
discussed in this section.

\section{Stochastic Master Equation}

In this section we turn to the quantum trajectory approach known from
quantum optics \citep{1986_PRA_Barchielli_Quntum_Trajectories,1986_Nagourney_Dehmelt_Quantum_Jumps,1987_Putterman_Porrati_PRA_Null_Measurement,1992_BraginskyKhalili_QuantumMeasurement,1993_Carmichael_Open_Quantum_Systems,1993_Cirac_Blatt_PRL_Quantum_Jumps_Ion_Trap,1995_Walls_Milburn_QuantumOpticsBook,Plenio1998Quantumjump,1999_PRL_Peil_Gabrielse_QND,2004_GardinerZoller_QuantumNoise,2004_10_Doherty_QND_FockMechanicalOscillator,2007_Haroche_Nature_Quantum_Jumps_of_light,gambetta2008qta}.
The stochastic master equation to be presented below allows us to
 model the backaction of the phase shift measurement and to produce
individual realizations of the measured phase shift signal. 

In the case of a cavity mode that decays much faster both than the
qubit decoherence rates ($\kappa_{cavity}\gg\gamma_{1},\gamma_{\phi})$
and the couplings to the cavity mode $\kappa_{cavity}\gg g_{i}\forall i\in\{0,1,..,N\}$,
it is possible to adiabatically eliminate the cavity mode from the
system and find for the stochastic master equation (in the interaction
picture) for the qubits alone

\begin{eqnarray}
\dot{\hat{\rho}} & = & \sum_{i=0}^{N}\gamma_{1}\left(\hat{\sigma_{i}}^{-}\hat{\rho}\hat{\sigma_{i}}^{+}-\frac{1}{2}\hat{\sigma_{i}}^{+}\hat{\sigma_{i}}^{-}\hat{\rho}-\frac{1}{2}\hat{\rho}\hat{\sigma_{i}}^{+}\hat{\sigma_{i}}^{-}\right)\nonumber \\
 & + & \sum_{i=0}^{N}\gamma^{\varphi}\left[2\hat{P}_{i}\hat{\rho}\hat{P}_{i}-\hat{P}_{i}\hat{\rho}-\hat{\rho}\hat{P}_{i}\right]\nonumber \\
 & - & 2\bar{\Gamma}\left[\hat{N},\left[\hat{N},\hat{\rho}\right]\right]\nonumber \\
 & - & \sqrt{4\bar{\Gamma}}\left(\hat{N}\hat{\rho}+\hat{\rho}\hat{N}-2\hat{\rho}\left\langle \hat{N}\right\rangle (t)\right)\xi(t).\,\,\,\,\,\label{eq:masterequation}\end{eqnarray}
Here $\bar{\Gamma}\equiv\frac{\bar{g}^{2}|\varepsilon|^{2}}{\kappa_{cavity}}$
is the measurement rate, $\gamma_{1}$ and $\gamma_{\varphi}$ are
the qubit relaxation and dephasing rates, $\hat{P}_{i}$ is the projector
onto the excited state of qubit $i$, and $|\varepsilon|^{2}$ is
the average photon number circulating inside the cavity mode. See
\citep{WisemanPhDThesis,2004_10_Doherty_QND_FockMechanicalOscillator}
for a detailed derivation, and \citep{helmer2007qze} for our recent
analysis of photon detection in circuit QED using the same approach.
The stochastic master equation is conditioned on the measured signal
\begin{equation}
X(t)\equiv\langle\hat{N}\rangle(t)+\frac{1}{4}\sqrt{\frac{1}{\bar{\Gamma}}}\xi(t),\label{eq:Signal}\end{equation}
where $\xi$ represents the fundamental, unavoidable vacuum noise
(with $\left\langle \xi(t)\xi(t')\right\rangle =\delta(t-t')$).
Physically, $X(t)$ is the appropriate (suitably normalized) quadrature
component of the electric field transmitted through the cavity, which
is proportional to the phase shift that indicates the multi-qubit
state. Experimentally, this signal would be measured in a homodyne
detection scheme. Note that, for a two-sided cavity, information is
contained both in the transmitted and the reflected signal, and we
have assumed that both parts of the signals are superimposed symmetrically
to extract the maximum possible information content \citep{helmer2007qze}.

\section{Examples of measurement-generated entangled states}

In this section we discuss the most relevant examples for the case
of two and three qubits in the cavity. More precisely we will show
that it is possible to generate Bell states, W states, and Greenberger-Horne-Zeilinger
states (GHZ-states). 

We will be able to observe that the measurement indeed first drives
the system to one of its attractor solutions (among them the desired
state) which are then stabilized by the measurement. The attractor
nature of the subspaces selected by the coupling vector $\overrightarrow{G}$
can be immediately understood from the structure of the stochastic
master equation (\ref{eq:masterequation}), by realizing that the
stochastic term and the measurement induced dephasing term both vanish
if the density matrix is in the desired state. Only relaxation and
dephasing can take the system out of this final state, and we will
discuss their influence later.

\subsection{Quantitative characterization}

In order to characterize the time-evolution during the measurement
process, we have plotted several quantities. We plot the phase shift
signal $X(t)$ and the excitation number $n_{i}$ in each qubit as
functions of time. To verify that we have indeed obtained the desired
state, we will compute the state fidelity $\digamma$ between the
density matrix from the simulation, $\rho_{sim}$, and the ideal state
density matrix $\sigma$, according to $\digamma\equiv{\rm tr}\left|\sqrt{\rho_{sim}}\sigma\sqrt{\rho_{sim}}\right|$.
Finally, the two-qubit entanglement between two qubits $A$ and $B$
will be measured by the log-negativity. Given the density matrix $\rho$
of the two qubits (after tracing out other qubits, if needed), this
is defined as $E_{N}(\rho)=\log_{2}\left\Vert \rho^{T_{A}}\right\Vert $.
Here $\rho^{T_{A}}$ is the partial transpose with respect to qubit
$A$, and $\left\Vert \mu\right\Vert ={\rm tr}[\sqrt{\mu^{\dagger}\mu}]$
is the trace norm. 

Before discussing the individual examples, we briefly point out the
general features. Looking at the results {[}e.g. in Fig.~(\ref{fig:2qubitWtraj})],
we find that in those cases where we end up in the right state the
fidelity as well as the log-negativity are 1. Furthermore, the state
is stabilized by the measurement, meaning that, due to the absence
of any non-vanishing terms in the master equation's right hand side,
it is frozen. We observe that the state is generated on a timescale
given by the measurement rate $\bar{\Gamma}^{-1}$. 

While discussing the examples we will also analyze plots that show
the probability density of various quantities evolving over time.
This point merits a brief discussion. The time-evolution of the distribution
for any simple quantum-mechanical observable can be immediately obtained
from the time-evolution of the \emph{average} density-matrix, i.e.
from the standard, non-stochastic master equation. In that case, simulating
a large number of stochastic trajectories and then averaging over
the results would be unnecessarily cumbersome. However, that argument
becomes void as soon as one considers signals that depend on the entire
pre-history of the trajectory. An important example is the time-averaged
cumulative phase-shift signal,

\begin{equation}
\bar{X}(t)\equiv\frac{1}{t}\int_{0}^{t}X(t')dt'\,.\end{equation}
This quantity has the advantage of tending towards a well-defined
limit in the course of a QND measurement, with the fluctuations around
that limiting value decreasing like $1/\sqrt{t}$. It is not possible
to obtain the distribution of $\bar{X}$ from the average density-matrix
$\rho$, and quantum jump trajectory simulations are needed.

Another example is represented by quantities that depend non-linearly
on the density matrix. In those cases, the average density matrix
is irrelevant since, obviously $\left\langle f(\rho)\right\rangle \neq f(\left\langle \rho\right\rangle )$
for a nonlinear function $f$. An important case is the entanglement
measure $E_{N}$. In fact, the average density matrix is never entangled
($E_{N}(\left\langle \rho\right\rangle )\equiv0$) for our examples.
Thus, it is indeed necessary to obtain $E_{N}$ for a large number
of trajectories in order to discuss its statistical behaviour and
plot the probability density.

\subsection{Bell states for two qubits - no decoherence}

In the case of two qubits and vanishing decoherence rates $\gamma_{1},\gamma_{\phi}=0$,
the generation of Bell states is straightforward \citep{2008arXiv0812.0218H}.
We imagine starting the experiment with all qubits in the ground state
$\ket{00}$ and applying a Hadamard gate ($\pi/2$ - $\sigma_{x}$
rotation) at some time $t_{0}$, which leaves the system in the product
state $\ket{\Psi_{0}}\equiv\prod_{\otimes}\frac{1}{\sqrt{2}}\left(\ket 0+\ket 1\right)=\frac{1}{2}\left(\ket{00}+\ket{01}+\ket{10}+\ket{11}\right)$.
We want to generate the Bell state

\begin{equation}
\ket{\Psi^{+}}\equiv\frac{1}{\sqrt{2}}\left(\ket{01}+\ket{10}\right),\end{equation}
which is the two-qubit version of a W-state. Clearly the amplitude
vector for this state is simply $ $$\sqrt{2}\overrightarrow{\alpha}=(0,1,1,0)^{T}$,
and the resulting Eq.~(\ref{eq:phase_shift_eq_general}) for the
couplings is given by $g_{1}=g_{2}$, thus $\overrightarrow{G}=(1,1)^{T}$.
The desired state will be generated with a success rate $\eta$ given
by \begin{equation}
\eta\equiv\left|\bra{\Psi^{+}}\Psi_{0}\rangle\right|^{2}=\frac{1}{2}\,,\end{equation}
meaning that the experiment will in 50\% of all runs end up in the
correct state (as confirmed by observation of the correct phase shift). 

\begin{figure}
\includegraphics[width=1\columnwidth]{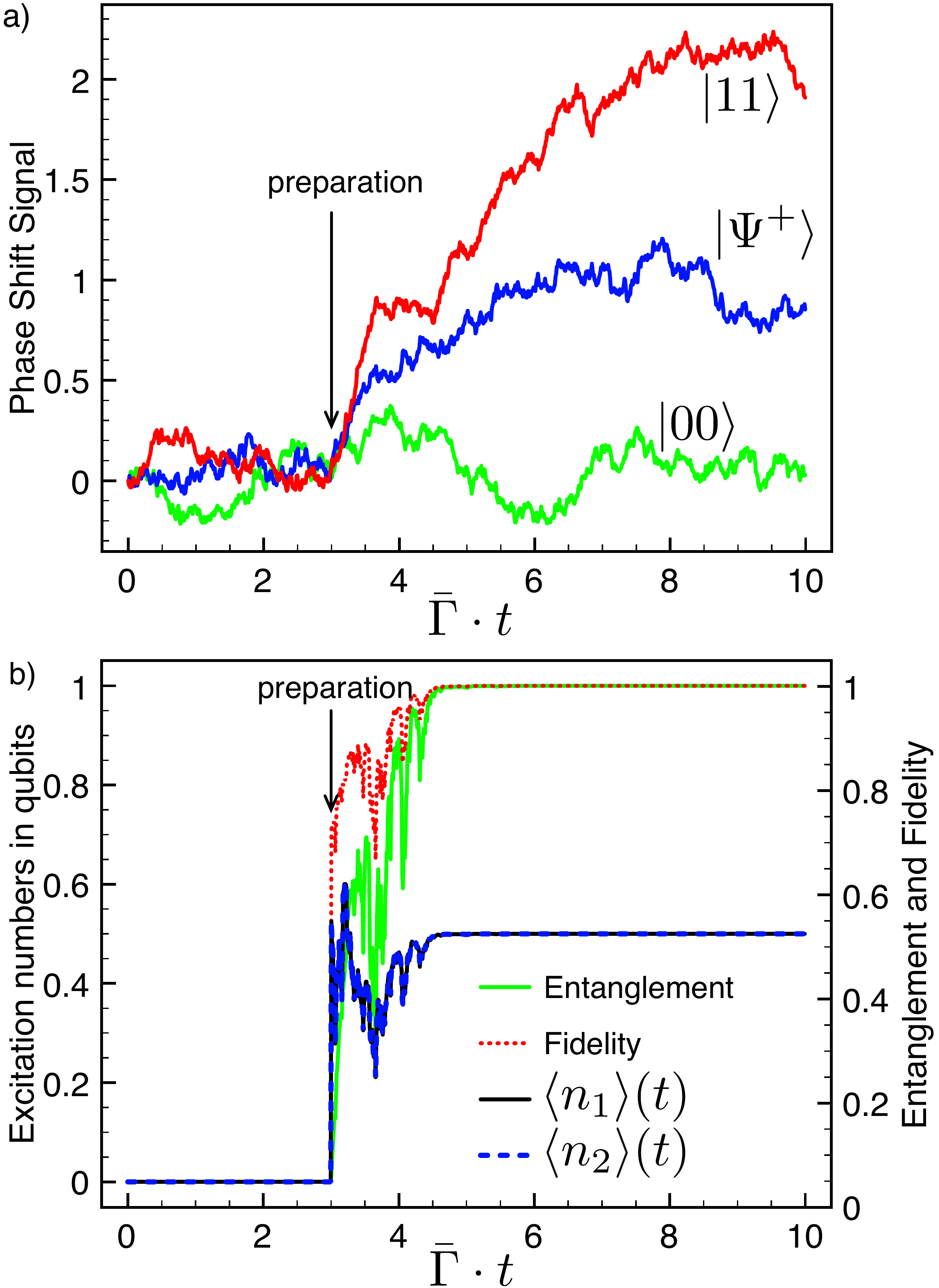}

\caption{(Color online) Generation of the Bell state $\ket{\Psi^{+}}=(\ket{10}+\ket{01})/\sqrt{2}$
(the two-qubit W-state): (a) Quantum trajectories illustrating the
different phase shift signal traces $X(t)$. Three traces have been
selected, corresponding to the possible outcomes of the measurement
given the same input state $\ket{\Psi_{0}}$. At time $\bar{\Gamma}t_{0}=3$,
Hadamard gates are applied to both qubits, starting from the ground
state. As in every real measurement of field quadratures, the signal
$X(t)$ is smoothed by doing a windowed average over a suitable time-span,
$\bar{\Gamma}\tau_{{\it avg}}=1.0$. Part (b) displays the excitation
numbers, state synthesis fidelity and the entanglement (log-negativity)
for the one trajectory of plot (a) that ended up in the desired state.}
\label{fig:2qubitWtraj}
\end{figure}

Likewise, for the Bell state vector $\ket{\Phi^{+}}\equiv\frac{1}{\sqrt{2}}\left(\ket{00}+\ket{11}\right)$, we
find for the amplitude vector $\sqrt{2}\overrightarrow{\alpha}=(1,0,0,1)^{T}$,
and for the characteristic equation for the couplings $g_{1}=-g_{2}$,which
is fulfilled by the choice of coupling vector $\overrightarrow{G}=(1,-1)^{T}$.
Note that in principle $\ket{\Phi^{+}}$ could also be generated by
first producing $\ket{\Psi^{+}}$ and then applying local unitary
operations, and the same is true for the two other Bell states, $\ket{\Psi^{-}}$
and $\ket{\Phi^{-}}$.

\begin{figure}
\includegraphics[width=1\columnwidth]{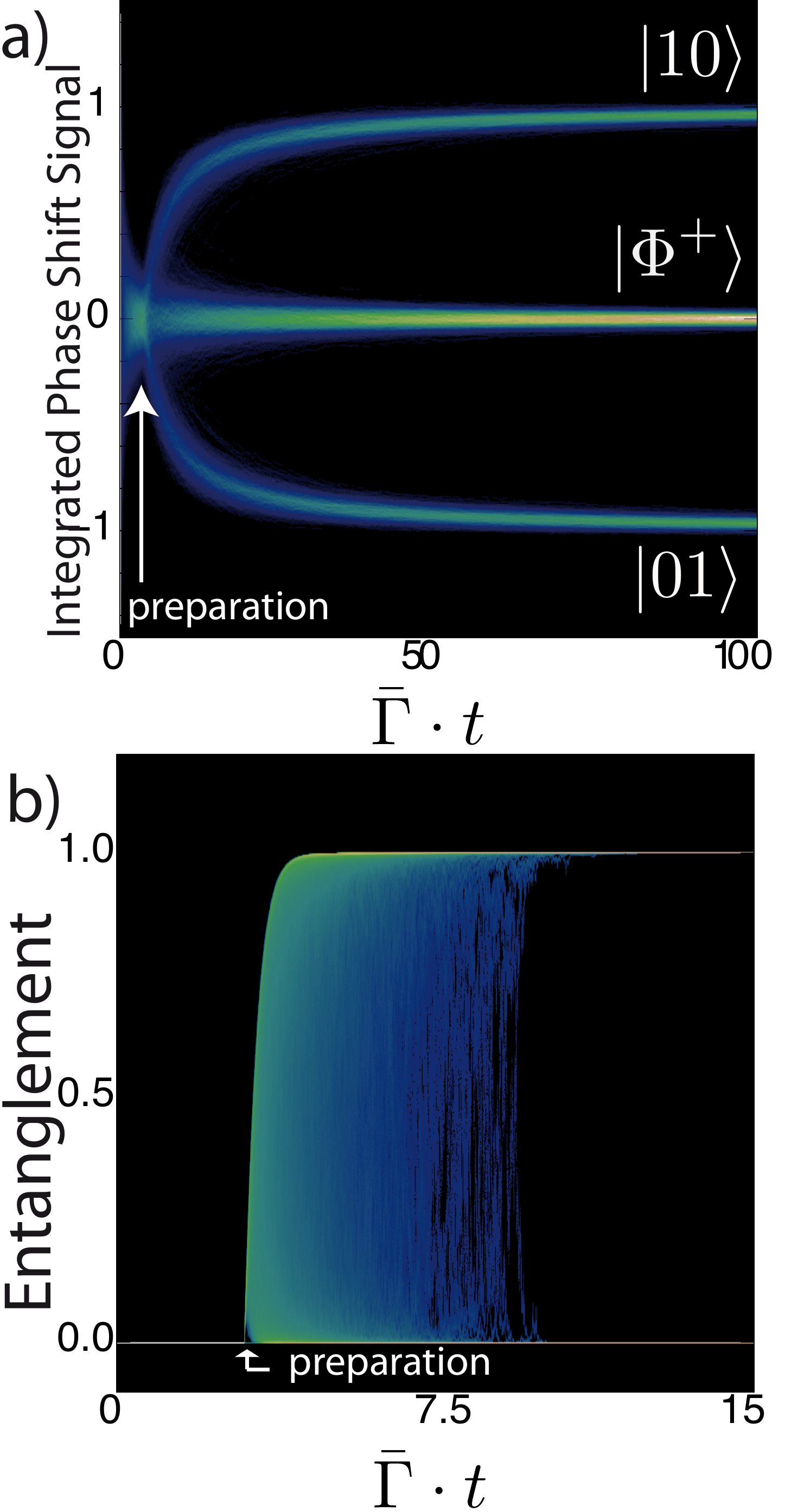}

\caption{(Color online) Generation of two-qubit Bell state $\ket{\Phi^{+}}=(\ket{00}+\ket{11})/\sqrt{2}$:
(a) Probability density of the integrated (cumulative) phase shift
signal $\bar{X}(t)=t^{-1}\int_{0}^{t}X(t')dt'$ from 6000 runs of
the simulation. At time $t_{0}$ Hadamard gates are applied to both
qubits. Part (b) displays the probability density of the entanglement
measure $E_{N}$, the log-negativity. Note that neither of these plots
can be obtained from the standard, non-stochastic master equation
(see main text), i.e. quantum jump trajectory simulations are essential.}
\label{fig:2qubitGHZtraj}
\end{figure}

Individual traces and probability density time evolutions for various
quantities are shown in Figs. \ref{fig:2qubitWtraj} and \ref{fig:2qubitGHZtraj},
respectively, for the two types of Bell states discussed here.

\subsection{Three qubits - no decoherence}

\subsubsection{Generation of W-states}

\begin{figure}
\includegraphics[width=1\columnwidth]{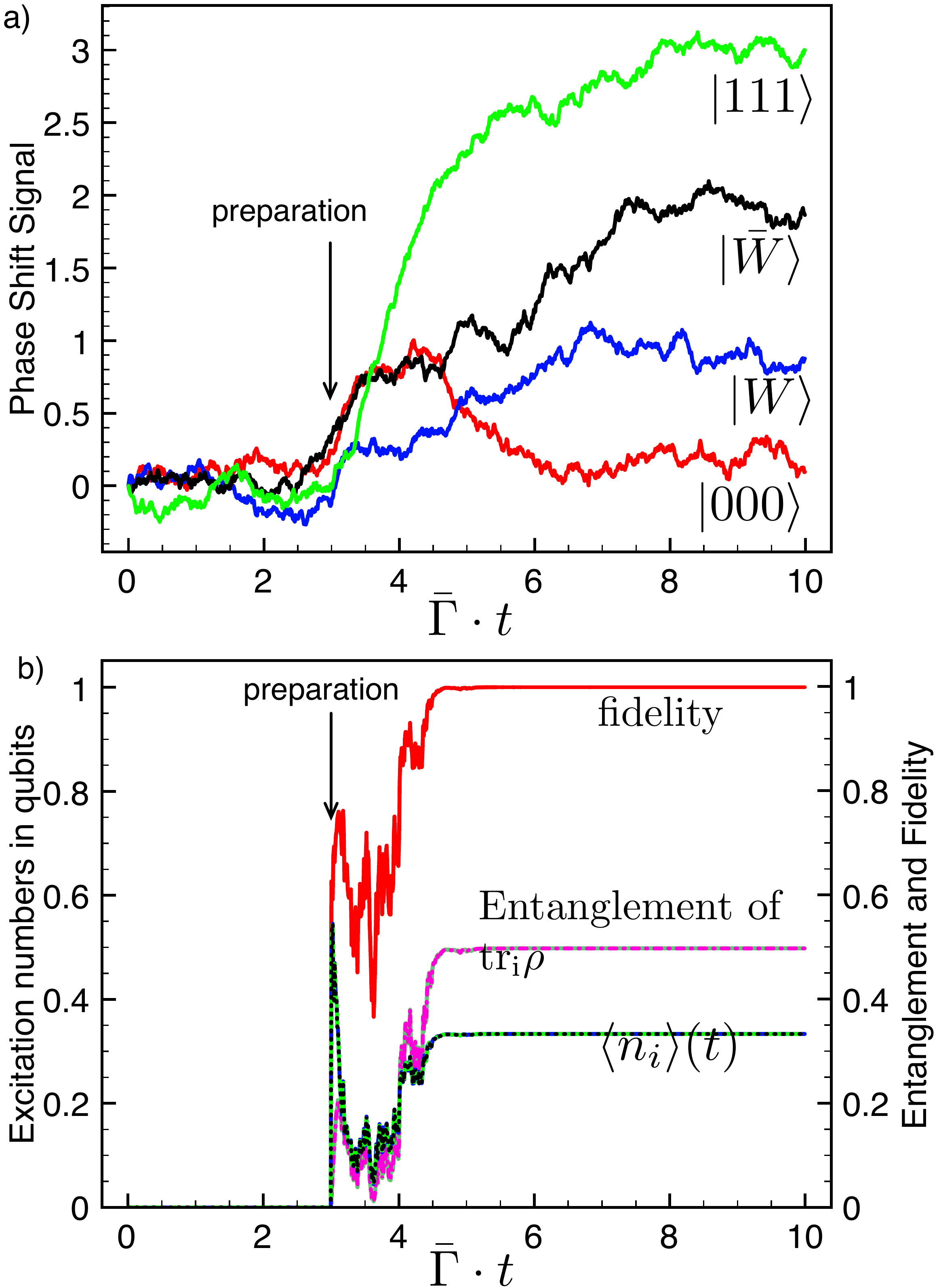}

\caption{(Color online) Generation of three qubit W-states: (a) Quantum trajectories
for the different states that can arise from the given input state
$\ket{\Psi_{0}}$. At time $\bar{\Gamma}t_{0}=3$, Hadamard gates
are applied to all qubits. Windowed averaging is performed as in Fig.~(\ref{fig:2qubitWtraj}).
Part (b) displays the excitation numbers, state synthesis fidelity
and the log-negativity for the one trajectory of plot (a) that ended
up in the desired W-state. Here $tr_{i}\rho$ denotes the partial
trace over qubit number $i$, and the resulting pairwise entanglement
happens to be the same for all choices of qubit pairs in this example.
Note that in the target state all pairs of qubits are mutually entangled
which is characteristic for the W-state and the reason for the robustness
of its entanglement compared to the GHZ state.}
\label{fig:3qubitWtraj}
\end{figure}

Similarly, for three qubits, the generation of W-states is straightforward
as well. We imagine starting the experiment with all qubits in the
ground state $\ket 0$, applying a Hadamard gate ($\pi/2$ - $\sigma_{x}$
rotation) at some time $t_{0}$, leaving the system in the state \begin{eqnarray*}
\ket{\Psi_{0}} & \equiv & \prod_{\otimes}\frac{1}{\sqrt{2}}\left(\ket 0+\ket 1\right)\\
 & = & \frac{1}{\sqrt{8}}\left[\ket{000}+\ket{001}+\ket{010}+\ket{011}\right.\\
 & + & \left.\ket{100}+\ket{101}+\ket{110}+\ket{111}\right]\end{eqnarray*}
 We aim to generate a W-state which for three qubits is given by

\[
\ket W\equiv\frac{1}{\sqrt{3}}\left(\ket{001}+\ket{010}+\ket{100}\right).\]
We find the corresponding amplitude vector $ $$\sqrt{3}\overrightarrow{\alpha}=(0,1,1,0,1,0,0,0)^{T},$
and the resulting equations for the couplings, $g_{1}=g_{2}=g_{3}$,
solved by equal couplings to all qubits, $\overrightarrow{G}=\left(\begin{array}{c}
1,1,1\end{array}\right)^{T}$. The W-state will be generated with a success rate $\eta$ given
by \[
\eta\equiv\left|\bra W\Psi_{0}\rangle\right|^{2}=\frac{3}{8}\,.\]
Note that with the same success rate the dual W state $ $\[
\ket{\bar{W}}\equiv\frac{1}{\sqrt{3}}\left(\ket{011}+\ket{110}+\ket{101}\right)\]
is generated (see Fig.~(\ref{fig:3qubitWtraj})). 

\subsubsection{Generation of GHZ-states}

\begin{figure}
\includegraphics[width=1\columnwidth]{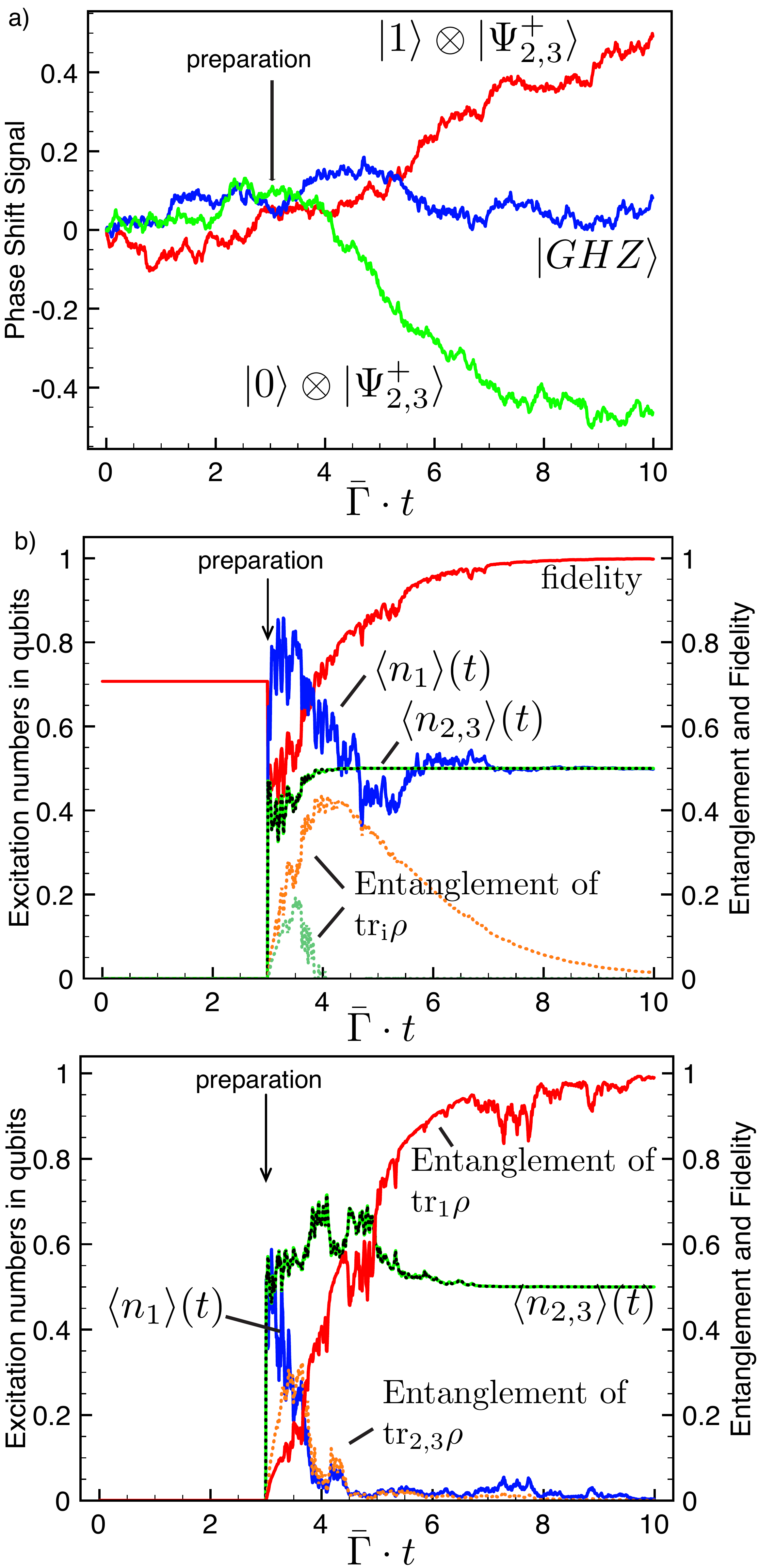}

\caption{(Color online) generation of three-qubit GHZ states: (a) Quantum trajectories
corresponding to the different states that can arise from the given
input state. At time $t_{0}$, Hadamard gates are applied to both
qubits. Note that among the unwanted outcomes there are two-qubit
$\ket{\Psi^{+}}$ Bell-states. These are actually generated with a
success rate of $\eta=3/4$ which is higher than in the original two-qubit
scheme. Part (b) displays the excitation numbers, state synthesis
fidelity and the log-negativity for all pairs of qubits for the trace
of part (a) that ended up in the desired GHZ-state. Note that once
the GHZ-state is reached, \emph{all }pairwise entanglement is lost.
This is a typical feature of GHZ states, which contain only genuine
three-particle entanglement. Part (c) shows the evolution for the
particular trajectory that reaches the Bell-state between qubits 2
and 3, which can be generated very efficiently as a byproduct using
this 3-qubit GHZ scheme.}
\label{fig:3qubitGHZtraj}
\end{figure}

Extending the two qubit EPR scheme to three qubits, we find for the
amplitude vector $\sqrt{3}\overrightarrow{\alpha}=\left(1,0,0,0,0,0,0,1\right)^{T}$,
and for the characteristic equation for the couplings in case of a
desired GHZ state as the target state $\ket{GHZ}\equiv\frac{1}{\sqrt{2}}\left(\ket{000}+\ket{111}\right)$:

\begin{eqnarray*}
0 & = & g_{1}+g_{2}+g_{3}\end{eqnarray*}
which is fulfilled, for example, by the choice of coupling vector
$\overrightarrow{G}=\left(1,-1/2,-1/2\right)^{T}$. The success rate
is $\eta=\frac{1}{4}$. Again we plot phase shift signal, excitation
numbers, log-negativity and fidelity to illustrate the correctness
of our considerations (see Fig.~(\ref{fig:3qubitGHZtraj})). Note
that due to the unequal couplings, the qubit excitations and pairwise
entanglement do depend on the qubit index, in contrast to all our
previous examples, where the couplings had been equal in magnitude.

It is noteworthy that this 3-qubit GHZ scheme yields a 75\% chance
of obtaining a W-state between qubits 2 and 3 as a byproduct. So this
might in fact be also considered an even more efficient scheme to
generate 2 qubit W-states than just with two qubits in the cavity.

\section{Effects of Decoherence}

We include decoherence into our model by considering the stochastic
master equation Eq.~(\ref{eq:masterequation}) with the Lindblad
decay and dephasing rates now different from zero. Assuming equal
rates for all the qubits, evidently entanglement will be on average
destroyed on a timescale set by $T_{2}=(\gamma_{1}/2+\gamma_{\phi})^{-1}$.
When considering experimentally reachable parameters, which we will
do further below, we will find that indeed the time needed to synthesize
states is orders of magnitude shorter compared to $T_{2}$. It is
thus clear that the simulation of the examples will look like above
with a weak decay of coherence superimposed on the trajectories. 

In contrast to the decay due to decoherence, the decay due to relaxation
(at a rate $\gamma_{1}$) is stochastic, in the sense that it leads
to sudden quantum jumps. This can be understood by considering that
the phase shift measurement stabilizes a certain subspace. Doing so,
certain configurations of diagonal elements in the density matrix
constitute attractors that compete with the exponential decay due
to $\gamma_{1}$. More formally speaking, the master equation is unravelled
with respect to the $\gamma_{1}$-process, but still an ensemble average
description of the pure dephasing physics. The result is that the
decay is stochastic when looking at single trajectories and the usual
exponential $\gamma_{1}$-decay is recovered when averaging over many
trajectories. Conversely, in a single trajectory the off-diagonal
elements decay on a timescale set by $T_{2}$, showing the following
behavior: As long as the relaxation jump process has not happened,
one observes a decay \emph{solely }due to pure dephasing (see Fig.~(\ref{fig:2qubitW_decoh-2})).
Once the relaxation process has happened, coherence and thus entanglement
are also lost completely.

\subsection{Example - two qubit Bell-states including dissipation}

To demonstrate the influence of decoherence and relaxation, we repeat
the example for a two qubit Bell-state, $\ket{\Psi^{+}}\equiv\frac{1}{\sqrt{2}}\left(\ket{01}+\ket{10}\right)$,
assuming comparatively low values of $\bar{\Gamma}/\gamma_{1}=10$
and $\bar{\Gamma}/\gamma_{\phi}=20$ to illustrate the effects and
make all the dynamics visible. Experimental ratios would be at least
about a factor 100 higher and thus the fidelity and lifetime are higher
in experiment than they appear from the following simulations. %
\begin{figure}
\includegraphics[width=1\columnwidth]{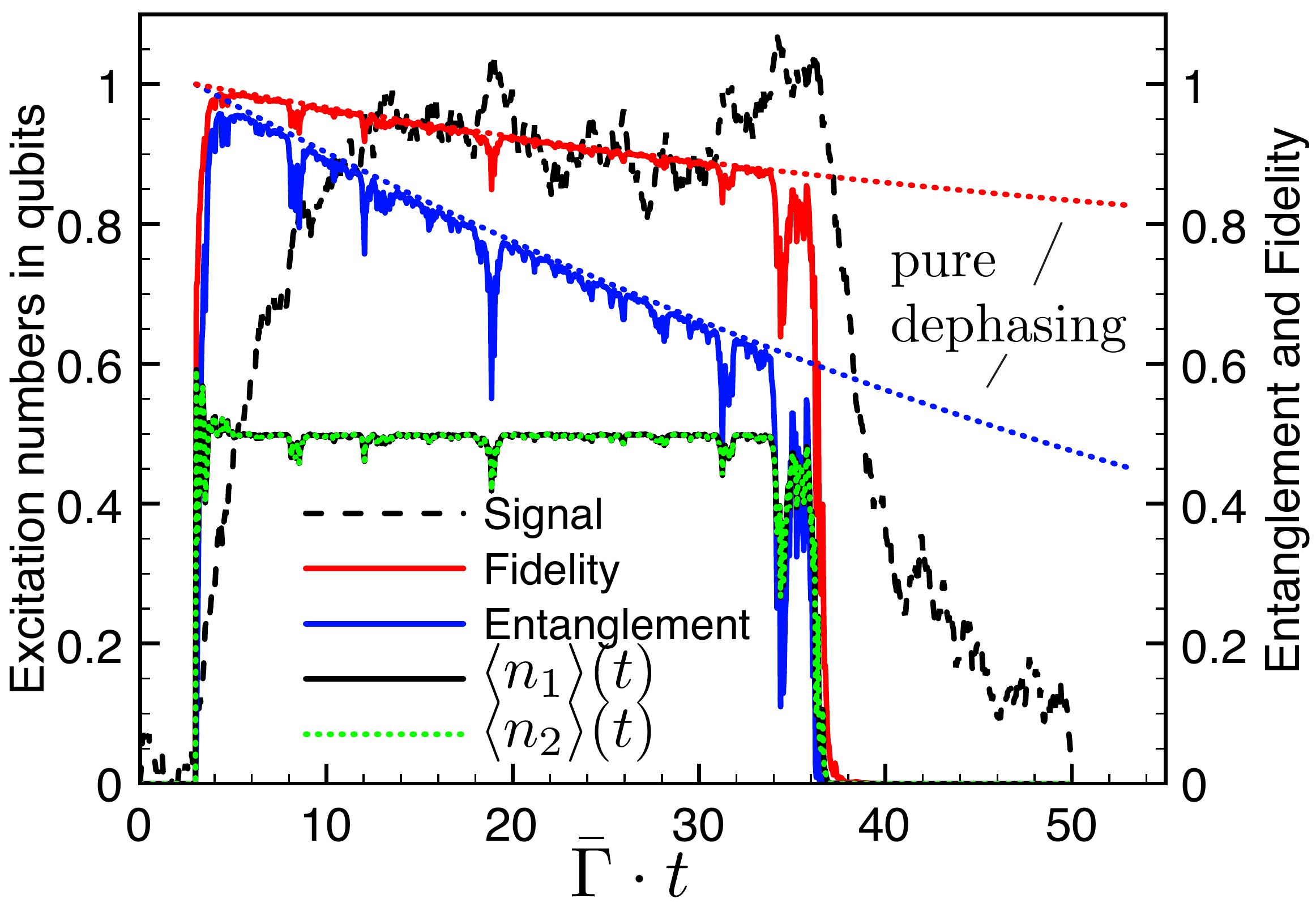}

\caption{(Color online) Effects of adding decoherence to the dynamics. The
situation is identical to the simulation of Fig.~(\ref{fig:2qubitWtraj}),
with the target state $\ket{\Psi^{+}}$, except for the added relaxation
rate $\gamma_{1}=0.01\cdot\bar{\Gamma}$ and pure dephasing rate $\gamma_{\phi}=0.02\cdot\Gamma$.
We can observe that the subspace of choice is stabilized before the
eventual decay due to relaxation. However, even before the sudden
jump due to relaxation, one observes a slow decay of the fidelity
and entanglement between the qubits, due to the pure dephasing rate
$\gamma_{\phi}$ (dashed lines). }
\label{fig:2qubitW_decoh-2}
\end{figure}

\begin{figure}
\includegraphics[width=1\columnwidth]{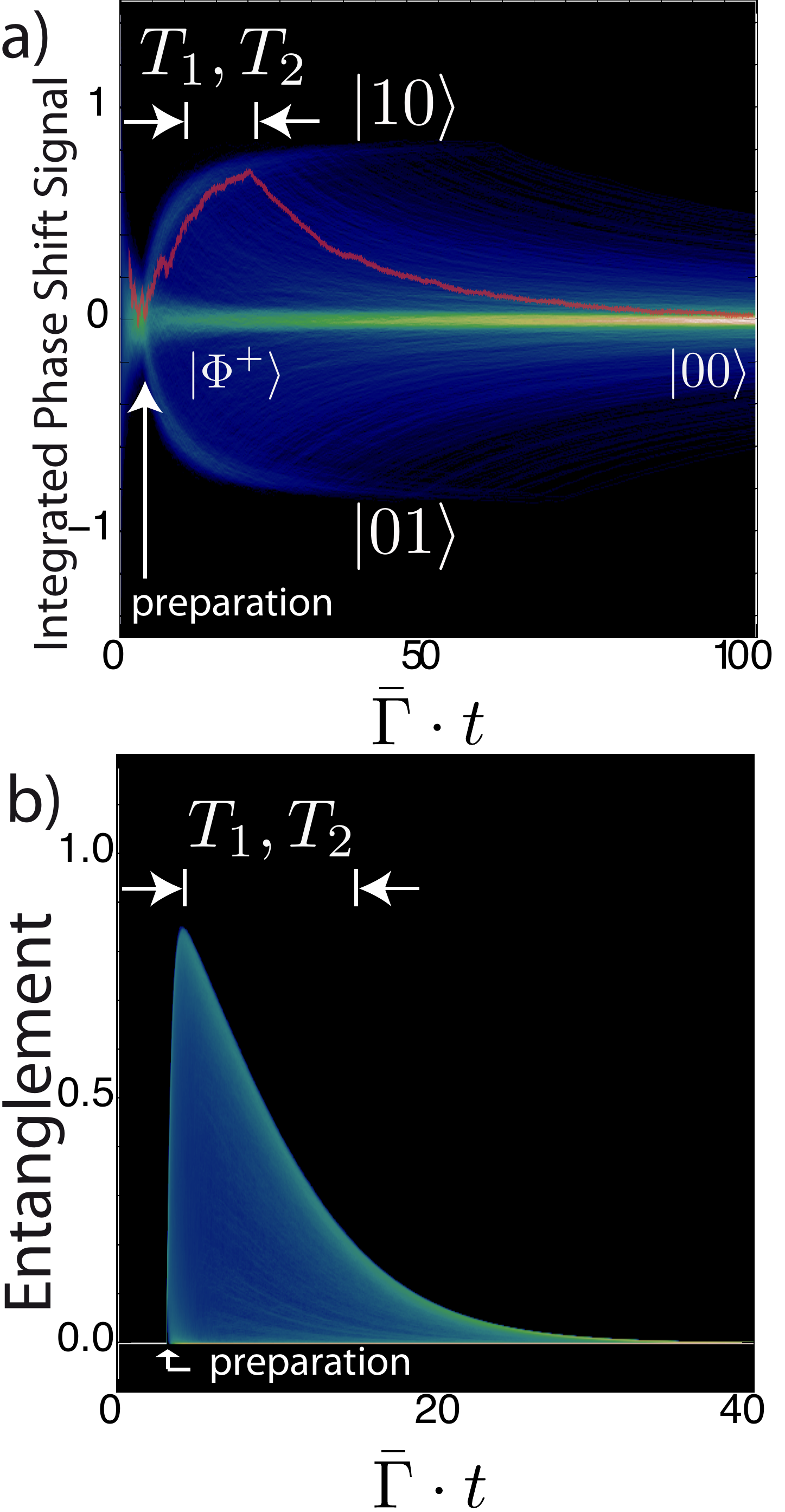}

\caption{(Color online) Effects of adding decoherence and relaxation to the
creation of the Bell state $\ket{\Phi^{+}}\equiv\frac{1}{\sqrt{2}}\left(\ket{00}+\ket{11}\right)$.
The situation is identical to the simulation of Fig.~(\ref{fig:2qubitGHZtraj}),
except for the added relaxation rate $\gamma_{1}=0.1\cdot\bar{\Gamma}$
and pure dephasing rate $\gamma_{\phi}=0.05\cdot\Gamma$. Part (a)
shows the probability density of the time-averaged (cumulative) phase
shift signal $\bar{X}(t)$ with an example trajectory shown in red.
Note the build-up of finite probability at finite signal values, before
relaxation back to zero phase shift, which represents the vacuum state
$\ket{00}$ at long times. Part (b) shows the probability density
of the entanglement (log-negativity). We can observe that the entanglement
is lost on a timescale given by $T_{2}=(\gamma_{1}/2+\gamma_{2})^{-1}$.
Note in particular the sharply defined, exponentially decaying envelope
that defines a \emph{strict} upper bound for the entanglement at any
given time. This is due to the pure dephasing. }
\label{fig:2qubitGHZ_decohdensity}
\end{figure}

We have plotted the time-evolution for the choice of couplings that
leads to the creation of a two qubit Bell-state $\ket{\Psi^{+}}\equiv\frac{1}{\sqrt{2}}\left(\ket{10}+\ket{01}\right)$.
The results are shown in Fig.~(\ref{fig:2qubitW_decoh-2}), which
should be compared against Fig.~(\ref{fig:2qubitWtraj}). Likewise,
we have considered the probability density for the time-averaged phase
shift signal and the entanglement measure, for the Bell state $\ket{\Phi^{+}}$,
see Fig.~\ref{fig:2qubitGHZ_decohdensity}. There, the strict upper
envelope for the entanglement is particularly noteworthy, corresponding
to the decay of coherence within the subspace selected by the measurement.

\section{Effect of imperfections due to parameter spread}

In order to prepare states in this way experimentally, one faces the
problem that it might not always be possible to fix important parameters
perfectly. If the scheme one has in mind in turn relies on exact matching
of parameters too much, one quickly ends up with a proposal that may
be interesting but not very realistic. We therefore examine the effects
on the fidelity and entanglement properties of this state synthesis
scheme in the presence of small deviations in the couplings of the
qubits to the cavity 

\begin{equation}
\overrightarrow{G}=\overrightarrow{G_{0}}+\left(\frac{\delta g_{i}}{\bar{g}}\right),\end{equation}
where $\overrightarrow{G_{0}}$ is the ideal coupling vector from
solving the characteristic equations Eq.~(\ref{eq:phase_shift_eq_general})
for the target state. $\delta g_{i}/\bar{g}$ are the deviations from
that ideal coupling for each qubit. Without loss of generality we
first look at the case where all couplings are equal to their ideal
value, except one which differs by $\delta g/\bar{g}$. We focus on
the stochastic term in the master equation Eq.~(\ref{eq:masterequation})
which is responsible for the projection onto a set of states, one
of which is our target state. Let us rewrite this term a bit by inserting
the definitions of $\bar{\Gamma}$ and $\hat{N}$: \begin{eqnarray}
\dot{\hat{\rho}}_{{\rm st}} & = & -\sqrt{4\bar{\Gamma}}\left(\hat{N}\hat{\rho}+\hat{\rho}\hat{N}-2\hat{\rho}\left\langle \hat{N}\right\rangle (t)\right)\xi(t)\nonumber \\
\nonumber \\ & = & -\sqrt{4\frac{\bar{g}^{2}|\alpha|^{2}}{\kappa_{cavity}}}\sum_{i=1}^{N}\frac{g_{i}}{\bar{g}}\left(\hat{n}_{i}\hat{\rho}+\hat{\rho}\hat{n}_{i}-2\hat{\rho}\left\langle \hat{n}_{i}\right\rangle (t)\right)\xi(t).\,\,\,\,\,\,\,\,\,\label{eq:stochastic_term_explicit}\end{eqnarray}

From this form of Eq.~(\ref{eq:stochastic_term_explicit}), it is
evident that due to the linearity in the couplings $g_{i}$, we can
pull out all deviating couplings into separate terms which have the
same form. This reads 

\begin{eqnarray*}
\dot{\hat{\rho}} & = & -\sqrt{4\bar{\Gamma}}\sum_{i=1}^{N}\frac{g_{i}^{(0)}}{\bar{g}}\left(\hat{n}_{i}\hat{\rho}+\hat{\rho}\hat{n}_{i}-2\hat{\rho}\left\langle \hat{n}_{i}\right\rangle (t)\right)\xi(t)\,\,\,\,\,\,\,\,\,\\
 & - & \sqrt{4\bar{\Gamma}}\sum_{i=1}^{N}\frac{\delta g_{i}}{\bar{g}}\left(\hat{n}_{i}\hat{\rho}+\hat{\rho}\hat{n_{i}}-2\hat{\rho}\left\langle \hat{n}_{i}\right\rangle (t)\right)\xi(t)\,\,\,\,\,\,\,\,\\
 & = & -\sqrt{4\bar{\Gamma}}\left(\hat{N}\hat{\rho}+\hat{\rho}\hat{N}-2\hat{\rho}\left\langle \hat{N}\right\rangle (t)\right)\xi(t)\,\,\,\,\,\,\,\,\\
 & - & \sum_{i=1}^{N}\sqrt{4\delta\Gamma_{i}}\left(\hat{n}_{i}\hat{\rho}+\hat{\rho}\hat{n}_{i}-2\hat{\rho}\left\langle \hat{n}_{i}\right\rangle (t)\right)\xi(t),\end{eqnarray*}
which means that in addition to the ideal behaviour captured by the
first term, each individual qubit with deviating coupling will be
projected on its ground or excited state on a timescale given by the
inverse of the individual measurement rate $\delta\Gamma_{i}\equiv\frac{|\varepsilon|^{2}\delta g_{i}^{2}}{\kappa_{{\rm cavity}}}$
(we have assumed positive $\delta g_{i}$ for simplicity; otherwise
the signs in the last line would change for those qubits with $\delta g_{i}<0$).
This has two consequences: The first consequence concerns the measured
phase shift: Instead of being equal for all the base kets that form
our target state, there will be deviations in the phase shift from
base ket to base ket. This means that we will be able to observe the
breakdown of the target state. Therefore, second, the lifetime of
the desired entangled state will now also be limited by the inverse
of the maximum of the individual measurement rates, in addition to
the effects of decoherence. 

In other words: As soon as we have gained enough signal to noise ratio
to discriminate the different base kets from each other (i.e. resolve
the different corresponding phase shifts), our target state will be
destroyed. 

\begin{figure}
\includegraphics[width=1\columnwidth]{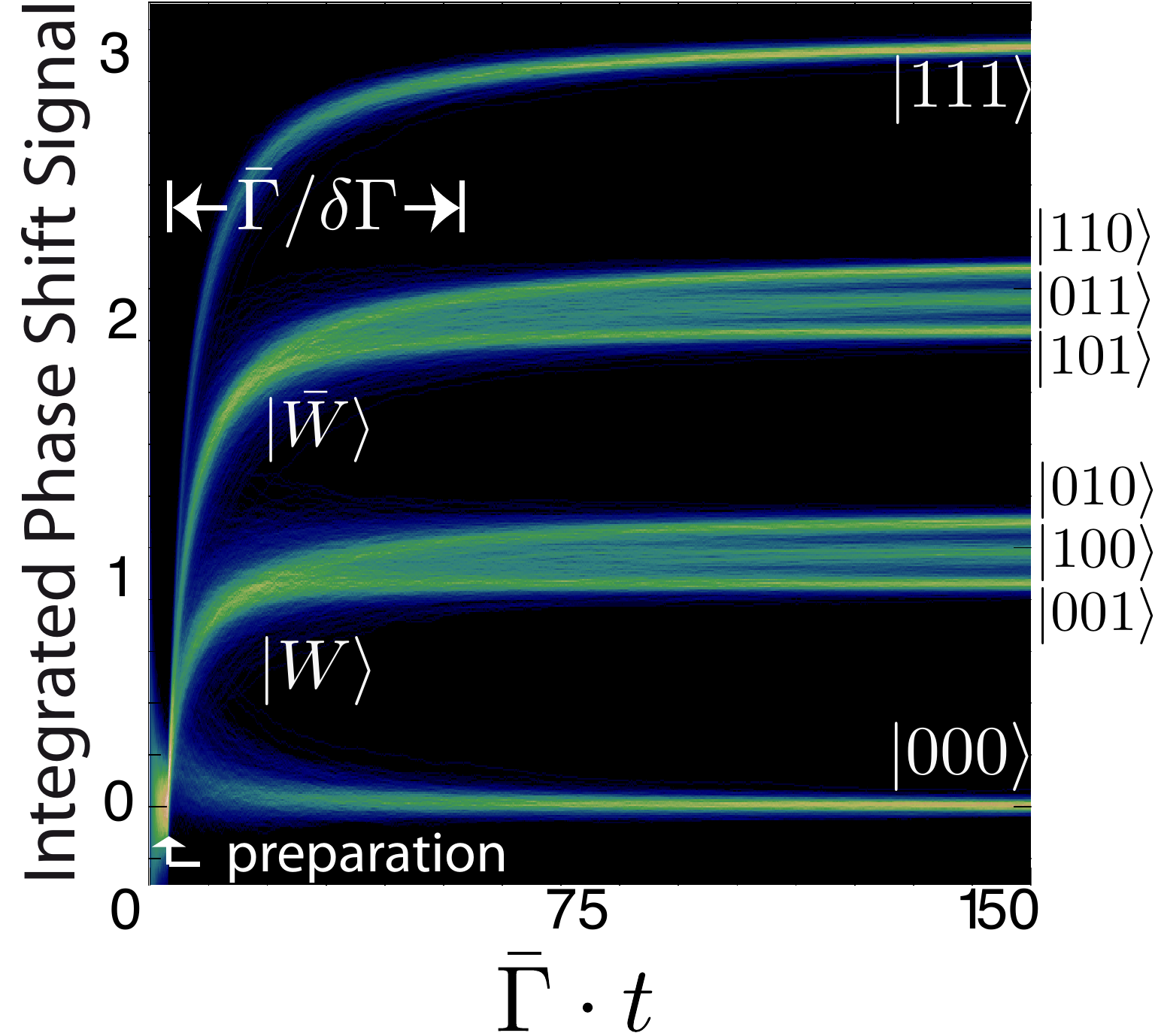}

\caption{(Color online) Probability density of the time-averaged cumulative
(integrated) phase shift signal $\bar{X}$ as an illustration of the
effect of parameter spread in the couplings. Apart from the deviation
in the couplings from the ideal values, the setup is identical to
the example in which we aimed for a three qubit W-state, as seen in
Fig.~\ref{fig:3qubitWtraj}. Hadamard gates are applied to all qubits
at time $t_{0}=3\cdot\bar{\Gamma}^{-1}$. During the following time
interval of length $\bar{\Gamma}^{-1}$ all trajectories are projected
onto the W-state, $\ket{\bar{W}}$ ,$\ket{000}$, or $\ket{111}$.
Meanwhile the competing projection on the individual number states
of the qubits becomes more pronounced and dominates the dynamics on
a time scale $\delta\Gamma_{2}^{-1}=50\bar{\Gamma}^{-1}$. This is
exactly the timescale on which we can be sure to identify all the
product base states by their phase shift values individually. }
\label{fig:parspread2qubitW}
\end{figure}

To illustrate this effect in a fairly drastic way, we choose an example
of three qubits with a W-state as a target state and the coupling
vector $ $$\overrightarrow{G}=(1,1,1)^{T}+(\sqrt{2}/10,0,-\sqrt{2}/10)^{T}$
. This yields an individual measurement rate for the second qubit
of $\delta\Gamma_{2}/\bar{\Gamma}=1/50$. Therefore, we expect the
target state and especially its entanglement properties to be destroyed
on a timescale of 50 times the preparation time $\bar{\Gamma}^{-1}$.
As we will argue in the following section, present experiments allow
a ratio $\bar{\Gamma}/\gamma_{decoh}=O(10^{4})$ which justifies to
ignore decoherence for the moment. The resulting simulation beautifully
confirms the expectations, see Fig.~\ref{fig:parspread2qubitW}.

We conclude that in order to observe the full dynamics of the system
one should strive for a regime where the condition \begin{equation}
\bar{\Gamma}\gg\delta\Gamma_{i}>\gamma_{1},\gamma_{\phi}\forall i\in1,...,N\label{eq:timescale_relation}\end{equation}
 is met. In the next section we will show that this is indeed possible
with present-day experimental parameters.

As a side remark we state that the situation of one coupling deviating
from the others is in principle already found when synthesizing GHZ
states for an odd number of qubits (e.g. 3) as examined in the previous
sections (see Fig.~(\ref{fig:3qubitGHZtraj}, a))). We had chosen
a coupling vector $\overrightarrow{G}=\left(1,-1/2,-1/2\right)^{T}$.
Here the larger magnitude of the coupling for the first qubit is responsible
for the generation of 2-qubit W-states. As a consequence, we can learn
about the state of the first qubit while we can still not distinguish
qubits 2 and 3 from each other. Following our previous reasoning in
this section, we find that the state of qubit 2 should be discerned
on a timescale set by $\left(\delta\Gamma_{2}/\bar{\Gamma}\right)^{-1}=4$
(in units of $\bar{\Gamma}^{-1}$), which matches the simulation results
shown in the previous section (see Fig.~\ref{fig:3qubitGHZtraj}a).

\section{Possibilities For Experimental Realization\label{sec:Possibilites-of-Experimental}}

Cavity QED setups in superconducting circuits \citep{2004_09_WallraffEtAl_MicrowaveCavity,2007_Yale_nature_Single_Photon_Source,1977_J_Math_Phys_Quantum_Zeno,2006_PRL_Johansson_96_127006}
have been used to implement ideas of quantum optics on the chip, and
are considered a promising candidate for scalable, fault tolerant
quantum computing (e.g. \citep{2007_Helmer_CavityGrid_arxiv}). Proposals
for generating and detecting non-classical photon states exist or
have been implemented \citep{2005_Storcz_Arxiv_On-chip_Fock_States,2006_05_Marquardt_PDC,2007_Yale_nature_Single_Photon_Source,2007_Yale_Nature_Photon_Number_Splitting,helmer2007qze}.

These experiments realize a Jaynes-Cummings coupling between qubit
and resonator of up to $2\pi\cdot100{\rm \, MHz}$, resonators with
frequencies of about $2\pi\cdot5{\rm GHz}$, and a large range of
resonator decay rates $\kappa$ between $1{\rm 0\, kHz}$ and $100{\rm \, MHz}$.
Given this parameter space and assuming a bare qubit coupling of $g_{0}\approx2\pi\cdot100\, MHz$,
detunings in the $GHz$ range, $\left|\epsilon\right|^{2}\approx10$
photons in the readout cavity, and a qubit decay rate $\gamma_{1}\approx0.6\, MHz$,
it is easily possible to reach values of $\bar{\Gamma}/\kappa\approx10^{4}$.
This gives ample time for the state synthesis before decoherence starts
playing a role.  

Furthermore, couplings can be adjusted with enough accuracy such that
the state generation is also not limited by this factor. We can examine
the sensitivity of the ratio $\delta\Gamma/\bar{\Gamma}$ to small
deviations in the parameters. From $\delta\Gamma\propto\delta g^{2}$
and $\delta g=\delta(g_{0}^{2}/\Delta)$, we find $\delta\Gamma/\bar{\Gamma}=[2\delta g_{0}/g_{0}-\delta\Delta/\Delta]^{2}$. Assuming
an uncertainty about the bare value of the coupling of the qubits
to the cavity and an uncertainty about the qubit detuning of about
$5\%$ each we find that $\delta\Gamma\sim0.05^{2}\bar{\Gamma}$.
Note that this value is obtained without even considering the possibility
of actively compensating for the spread in the couplings by suitably
adjusting the detuning. This hints that under presently available
optimal experimental conditions, the infidelity due to parameter spread
becomes visible only long after the system has been severely decohered
anyway. However, one can always intentionally choose parameters such
that Eq.~(\ref{eq:timescale_relation}) is fulfilled and the full
dynamics discussed here can be experimentally observed, including
the ultimate measurement-induced decay of the temporarily produced
entangled state.

The main challenging step to be taken experimentally before realizing
this scheme in the lab is to operate in the single-shot qubit readout
limit. This has been demontrated very recently by the Saclay group
using a Josephson bifurcation amplifier setup \citep{2008_FMallet_Single_Shot_priv_vomm}.

\section{Conclusions}

We have analyzed a very general experimentally directly relevant way
to generate entangled multi-qubit states using a dispersive phase
shift measurement of the collective state of several qubits inside
a cavity. We have given criteria for the possibility to synthesize
a given target state and studied the most relevant examples of Bell-states
as well as W- and GHZ-states for two and three qubits. We have also
discussed, and analyzed by extensive numerical simulations, the two
major sources of imperfections in this setup, namely decoherence and
parameter spread. Finally, we have compared with presently reachable
experimental parameters and conclude that this scheme could soon be
tested in the laboratory.

\section{Acknowledgements }

We thank Jens Siewert for enlightening discussions. Support from the
SFB 631, NIM, and the Emmy-Noether program (F.M.) of the DFG, as well
as EuroSQIP, are gratefully acknowledged.

\bibliographystyle{apsrev}
\bibliography{SiPh}

\end{document}